\documentclass[10pt]{article}
\usepackage{lscape} 

\usepackage{comment} 
\def \ccomma{\raise 2pt\hbox{,}} 
\def \D {\hbox{d}}

\def \mod#1{\vert #1 \vert}
\def \barQ {\overline{Q}}
\def \barU {\overline{U}}
\def \barV {\overline{V}}
\def \barW {\overline{W}}
\def \bfp {{\bf p}}
\def \bfq {{\bf q}}
\def \bfu {{\bf u}}
\def \bfE {{\bf E}}
\def \bfv {{\bf v}}
\def \vg {v_g}
\def\genEx {E_x}   
\def\genEy {E_y}   
\def\genA  {A}     
\def\gVsix {V_6}   
\def\gWsix {W_6}   

\def\gVfou {V_4}   
\def\genGx {G_x}   
\def\genGy {G_y}   
\def\genHx {H_x}   
\def\genHy {H_y}   
\def\gBone {B_{1}} 
\def\gBtwo {B_{2}} 
\def\genex {e_x}   
\def\geney {e_y}   

\def\gWfou {W_4}   
\def\gVthr {V_3}   

\begin{document}

\title{Analytic study of a coupled Kerr-SBS system}

\author{Robert Conte${}^{1,2}$\footnote{Corresponding author.} and Maria Luz Gandarias${}^{3}$
{}\\
\\ 1. LRC MESO, 
      CEA-DAM-DIF, F-91297 Arpajon, France
\\2. Department of Mathematics, The University of Hong Kong,
\\ Pokfulam Road, Hong Kong.
\\    E-mailRobert.Conte@cea.fr 
{}\\
\\ 3. Departamento de Matematicas
\\ Universidad de C\'adiz
\\ Casa postale 40
\\ E--11510 Puerto Real, C\'adiz, Spain.
\\ E-mail:  MariaLuz.Gandarias@uca.es
}

\maketitle

\hfill 

{\vglue -10.0 truemm}
{\vskip -10.0 truemm}

\begin{abstract}
In order to describe the coupling between the Kerr nonlinearity
and the stimulated Brillouin scattering,
Mauger \textit{et al.} recently proposed a system of partial differential equations
in three complex amplitudes.
We perform here its analytic study by two methods. 
The first method is to investigate the structure of singularities,
in order to possibly find closed form singlevalued solutions obeying this structure.
The second method is to look at the infinitesimal symmetries of the system
in order to build reductions 
to a lesser number of independent variables.
Our overall conclusion is that the structure of singularities is too intricate
to obtain closed form solutions by the usual methods.
One of our results is the proof of the nonexistence of traveling waves.

\end{abstract}





\baselineskip=12truept 


\tableofcontents


\section{The coupled Kerr-SBS system}

The coupling between Kerr effect and stimulated Brillouin scattering \cite{AgrawalBook}
can be described
by three complex partial differential equations (PDE) in three complex amplitudes $U_1,U_2,Q$
depending on four independent variables $x,y,t,z$
\cite[Eqs.~(7)--(9)]{MBS2011}
\begin{eqnarray}
& &
\left\lbrace
\begin{array}{ll}
\displaystyle{
\phantom{+}
 i (U_{1,z} + \vg U_{1,t}) +\frac{U_{1,xx}+U_{1,yy}}{2 k_0}
 + b \left(\mod{U_1}^2+2 \mod{U_2}^2 \right) U_1 + i \frac{g}{2} Q U_2=0,
}\\ \displaystyle{
-i (U_{2,z} - \vg U_{2,t}) +\frac{U_{2,xx}+U_{2,yy}}{2 k_0}
 + b \left(\mod{U_2}^2+2 \mod{U_1}^2 \right) U_2 - i \frac{g}{2}\barQ U_1=0,
}\\ \displaystyle{
 \tau Q_t +Q - U_1 \barU_2=0,
}
\end{array}
\right.
\label{eqSBS466} 
\end{eqnarray}
in which $\vg,k_0,b,g,\tau$ are real constants.
We adopt the notation of nonlinear optics, in which the time $t$ and the
longitudinal coordinate $z$ are exchanged as compared to mathematical physics.

Although we will focus on the generic case $g \tau\not=0$,
we will also consider the two nongeneric cases $g \tau=0$,
for which the system is only four-dimensional,
\begin{eqnarray}
& & g \tau=0:
\left\lbrace
\begin{array}{ll}
\displaystyle{
\phantom{+}
 i (U_{1,z} + \vg U_{1,t}) +\frac{U_{1,xx}+U_{1,yy}}{2 k_0}
 + \left(b \mod{U_1}^2+\left(2 b + i \frac{g}{2}\right) \mod{U_2}^2 \right) U_1 =0,
}\\ \displaystyle{
-i (U_{2,z} - \vg U_{2,t}) +\frac{U_{2,xx}+U_{2,yy}}{2 k_0}
 + \left(b \mod{U_2}^2+\left(2 b - i \frac{g}{2}\right) \mod{U_1}^2 \right) U_2=0.
}
\end{array}
\right.
\label{eqSBS444} 
\end{eqnarray}

At present time, no solution is known to the generic system (\ref{eqSBS466}) ($\vg b g \tau \not=0$).
The goal of this work is to look for possible closed form solutions 
by two methods: singularity analysis, infinitesimal symmetries.

A prerequisite to the search of closed form solutions
is to investigate the singularity structure of the system,
this is done in section \ref{sectionPtest},
and this results in a triangular system of five PDEs to be obeyed
in order for a closed form solution to exist.

In section \ref{sectionOne-family},
we look for the simplest class of possible closed form solutions,
in which $U_1, U_2, Q$ could have shock profiles.
We find that, at least for the radial reduction $(U_j,Q)=f(x^2+y^2,z,t)$,
such a solution does not exist.

In section \ref{sectionClassicalSymmetries},
we apply the classical Lie method,
derive the Lie algebra,
compute the commutator table and the adjoint table \cite{OlverBook}.

Finally, in section \ref{sectionClassicalReductions},
we define a few reductions to a lesser number of independent variables.

\section{Singularity analysis}
\label{sectionPtest}

There exists only one limiting case in which the system (\ref{eqSBS466}) is integrable,
this is its degeneracy to the nonlinear Schr\"odinger equation
$U_1=U_2,g=0,\partial_z=0, c_1 \partial_x + c_2 \partial_y=0, (c_1,c_2)\not=(0,0)$.
Let us prove that, except for this limiting case, the system (\ref{eqSBS466}) is always nonintegrable,
in the sense that it always admits a multivalued behaviour around a singularity
which depends on the initial conditions (i.e.~what is called a movable singularity).
It is convenient to denote the list of dependent variables
$(U_1,\barU_1,U_2,\barU_2,Q,\barQ)$ as the six-dimensional vector $\bfu$.

A necessary condition for the system (\ref{eqSBS466}) 
to display a  
singlevalued behaviour of the general solution around any movable singularity
(i.e.~what is known as the Painlev\'e property \cite{CMBook})
is that all possible Laurent series locally representing the general solution
near a movable singular manifold $\varphi(x,y,z,t)-\varphi_0=0$,
\begin{eqnarray}
& &
\bfu =\sum_{j=0}^{+\infty} \bfu_j \chi^{j+\bfp},
\label{eqLaurent} 
\end{eqnarray}
indeed exist.
In the above series, the expansion variable $\chi$ vanishes when $\varphi(x,y,z,t)-\varphi_0 \to 0$.
This classical but technical computation \cite{CMBook} generates several necessary conditions,
the main ones being the following.

\begin{enumerate}
\item
At least one of
the six components of the leading power $\bfp$ must not be a positive integer
(so that $\chi=0$ is indeed a singularity).

\item
The Fuchs indices of the linearized system near the solution (\ref{eqLaurent}) must all be integer
(whatever be their sign).

\item
For any Fuchs index $j\ge 1$, the (affine) recurrence relation for $\bfu_j$
must admit a solution,
i.e.~no logarithms are allowed to enter the expansion,
and this requires some conditions (no-logarithm conditions, in short no-log conditions) to be obeyed.
\end{enumerate}

\subsection{Generic case $g \tau \not=0$ }
\label{sectionPtestgeneric}

There exists a dominant behaviour in which all six complex fields have simple poles
($\chi$ is here chosen as $\chi=\varphi(x,y,z,t)-\varphi_0$),
\begin{eqnarray}
& &
\left\lbrace
\begin{array}{ll}
\displaystyle{
    U_1 \sim M e^{ i a_1} \chi^{-1},\
\barU_1 \sim M e^{-i a_1} \chi^{-1},\
    U_2 \sim M e^{ i a_2} \chi^{-1},\
\barU_2 \sim M e^{-i a_2} \chi^{-1},\
}\\ \displaystyle{
    Q   \sim N e^{i a_1-i a_2} \chi^{-1},\
\barQ   \sim N e^{i a_2-i a_1} \chi^{-1},\
}\\ \displaystyle{
M^2=-\frac{\varphi_x^2+\varphi_y^2}{3 k_0 b},\
N=   \frac{\varphi_x^2+\varphi_y^2}{3 k_0 b \tau \varphi_t},\
(\varphi_x^2+\varphi_y^2) \varphi_t \not=0.
}
\end{array}
\right.
\label{eqSBS466gtau} 
\end{eqnarray}
and the two phases $a_1,a_2$ are arbitrary functions of $(x,y,z,t)$.
These two sets of values for the moduli $(M,N)$
define two families of movable singularities.
The Fuchs indices of each family are equal to
\begin{eqnarray}
& &
-1,0,0,1,1,3,3,4,\frac{3}{2}+\frac{\sqrt{11}}{2 \sqrt{3}},
                 \frac{3}{2}-\frac{\sqrt{11}}{2 \sqrt{3}},
\end{eqnarray}
and the two irrational indices prove the nonintegrability of the system.
This however does not yet rule out possible singlevalued solutions.

Each of the five indices $1,1,3,3,4$ generates one necessary condition 
for the Laurent series (\ref{eqLaurent}) to exist.
If they are all obeyed, the Laurent series
depends on the eight arbitrary 
functions
\begin{eqnarray}
& &
\varphi,a_1,a_2,Q_1,\barQ_1,U_{1,3}-\barU_{1,3},U_{2,3}-\barU_{2,3},
U_{1,4}+\barU_{1,4}+U_{2,4}+\barU_{2,4},
\label{eqSBS466gtau_arblist} 
\end{eqnarray}
associated to the respective Fuchs indices $-1,0,0,1,1,3,3,4$.
The five no-log conditions define a triangular system 
in the first five functions of this list, 
whose structure is
($P$ denotes a polynomial of all its arguments, having degree one in its first argument, 
$D^k f$ is the set of all derivatives of $f$ of order $k$,
and $D^{k:l}$ is a range of such derivatives),
\begin{eqnarray}
& &
\left\lbrace
\begin{array}{ll}
\displaystyle{
Q_{1,a}\equiv P(D^2 \varphi, D \varphi)=0,\ 
}\\ \displaystyle{
Q_{1,b}\equiv P(D (a_1-a_2), D \varphi)=0,\
}\\ \displaystyle{ 
Q_{3,a}\equiv P(D^2 (a_1+a_2), D^{3:1} \varphi)=0,\
}\\ \displaystyle{
Q_{3,b}\equiv P(D (Q_1+\barQ_1), D^2 (a_1-a_2), D^{3:1} \varphi)=0,\
}\\ \displaystyle{
Q_{4}\ \ \equiv P(D^2 (Q_1-\barQ_1), D (Q_1+\barQ_1), D^3 (a_1+a_2), D^3 (a_1-a_2), D^{5:1} \varphi)=0.
}
\end{array}
\right.
\label{eqSBS466gtau_nolog} 
\end{eqnarray}

Unless these five conditions are all obeyed,
no singlevalued particular solution exists,
therefore one must find at least one particular solution of this set of five conditions.

The two necessary conditions at Fuchs index $1$ are 
\begin{eqnarray}
& &
Q_{1,a}\equiv 
  3 \left(\varphi_x^2+\varphi_y^2\right)^2 (\tau^{-1} \varphi_t - \varphi_{tt})
 +6 \left(\varphi_x^2+\varphi_y^2\right)   \varphi_t (\varphi_x \varphi_{xt} +\varphi_y \varphi_{yt})
\nonumber \\ & & \phantom{123456}
 + \varphi_t^2 (\varphi_x^2 (3 \varphi_{xx}+\varphi_{yy})
               +\varphi_y^2 (3 \varphi_{yy}+\varphi_{xx}) + 4 \varphi_x \varphi_y \varphi_{xy})
=0,
\label{eqSBS466gtau_Q1a} 
\\ & &
Q_{1,b}\equiv 
\left[\varphi_x \partial_x +\varphi_y \partial_y 
      -\frac{\varphi_x^2+\varphi_y^2}{\varphi_t} \partial_t  
\right]\left(a_1-a_2+\frac{g t}{3 b \tau}\right) + 2 k_0 \varphi_z=0.
\label{eqSBS466gtau_Q1b} 
\end{eqnarray} 

The first condition admits no solution in the class
$\varphi=\Phi(\xi)$ with $\xi=k_x x + k_y+ k_z z +k_t t$
and $k_\alpha$ constants,
therefore
the generic ($g \tau\not=0$) Kerr-SBS system (\ref{eqSBS466}) 
admits no singlevalued travelling wave solution
(it does not admit plane waves either),
and numerical studies \cite{MBS2011} indeed confirm this result.

This first condition (\ref{eqSBS466gtau_Q1a}) is a quasilinear PDE of the Monge-Amp\`ere type.
Therefore, after switching to polar coordinates,
\begin{eqnarray}
& &
 (x,y) \to (\rho,\theta):\
 x=\rho \cos \theta,\ y=\rho \sin \theta,\
\label{eqpolar} 
\end{eqnarray}  
one follows the classical procedure of Goursat \cite[\S 24 p.~44, 1st English edition]{GoursatCours},
and performs a hodograph transformation like
\begin{eqnarray}
& &
\varphi(\rho,\theta,z,t) \to T(\varphi,\rho,\theta,z).
\end{eqnarray}  
This maps the two PDEs (\ref{eqSBS466gtau_Q1a})--(\ref{eqSBS466gtau_Q1b}) to 
an equivalent system,
which is even shorter when written for 
$T(\varphi,R,\theta,z)$ with $T=e^{t/\tau},R=\rho^{2/3}$,
\begin{eqnarray}
& & {\hskip -12.0truemm}
Q_{1,a}\equiv 
  9     T_\theta^2 T_R             T_{\varphi \varphi}
+ 4 R^2            T_R T_\varphi^2 T_{R       R}
+ 3                T_R T_\varphi^2 T_{\theta  \theta}
- 6     T_\theta^2     T_\varphi   T_{\varphi R}
-12     T_\theta   T_R T_\varphi   T_{\varphi \theta}
+ 6     T_\theta       T_\varphi^2 T_{R       \theta}
=0,
\label{eqSBS466gtau_Q1ahodo} 
\\ & & {\hskip -12.0truemm}
Q_{1,b}\equiv 
b T \left[ 6 T_R \partial_R 
+\frac{27}{2 R^2}
 \left(T_\theta \partial_\theta -\frac{T_\theta^2}{T_\varphi} \partial_\varphi\right)
\right] (a_1-a_2)
+ 2 g T_R^2 
+ 27 b k_0 R T T_z
=0.
\label{eqSBS466gtau_Q1bhodo} 
\end{eqnarray}
Any solution is acceptable provided it fulfills the condition 
\begin{eqnarray}
& & 
(R^2 T_R^2+T_\theta^2) T_\varphi \not=0,\
\end{eqnarray} 
inherited from the condition
$(\varphi_x^2+\varphi_y^2) \varphi_t \not=0$, see (\ref{eqSBS466gtau}),
\begin{eqnarray}
& & 
\varphi_x^2+\varphi_y^2=\frac{4}{9 R^3 T_\varphi^2} (R^2 T_R^2+T_\theta^2),\
\varphi_t =\frac{T}{\tau T_\varphi}.
\end{eqnarray} 

Two particular solutions of $Q_{1,a}=0$ are easy to obtain,
they are respectively defined by $T_R=0$ (azimutal reduction)
and $(T_\theta=0,T_{RR}=0)$ (radial reduction),
but only the radial reduction allows one to also integrate $Q_{1,b}=0$,
thus implicitly defining a particular solution of 
(\ref{eqSBS466gtau_Q1a})--(\ref{eqSBS466gtau_Q1b}) 
in terms of three arbitrary functions $G_0,G_1,G_2$ of two variables,
\begin{eqnarray}
& &
\partial_\theta=0:\ 
e^{t/\tau}=G_1(\varphi,z) (R +G_0(\varphi,z)),\
a_1-a_2= -\frac{g}{3 b} \left[\frac{t}{\tau} -G_2(\varphi,z)\right].
\label{eqSBS466gtau_Q1aQ1breducsol} 
\end{eqnarray}  

The next condition $Q_{3,a}=0$ to be solved in the triangular system (\ref{eqSBS466gtau_nolog})
is a second order linear PDE for $a_1+a_2$,
and, for the values (\ref{eqSBS466gtau_Q1aQ1breducsol}),
we could not find at least one solution.
 
\subsection{Case $g\not=0$ and $\tau=0$}
\label{sectionPtesttau0}

When $g\not=0$ and $\tau=0$,
the system (\ref{eqSBS444}) is made of two coupled complex Ginzburg-Landau equations
in 2+1 dimensions,
and, following the analysis made in \cite{CM2000b},
it admits a dominant behaviour ($\chi$ again denotes $\varphi(x,y,z,t)-\varphi_0$),
\begin{eqnarray}
& &
\left\lbrace
\begin{array}{ll}
\displaystyle{
    U_1 \sim M_1 e^{ i a_1} \chi^{-1+i \alpha},\
\barU_1 \sim M_1 e^{-i a_1} \chi^{-1-i \alpha},\
    U_2 \sim M_2 e^{ i a_2} \chi^{-1+i \beta},\
\barU_2 \sim M_2 e^{-i a_2} \chi^{-1-i \beta},\
}\\ \displaystyle{ 
M_1^2=\frac{(4  \beta^2-2 \alpha^2-4)b-3  \beta g}{12 b^2+g^2} \frac{\varphi_x^2+\varphi_y^2}{k_0},\
M_2^2=\frac{(4 \alpha^2-2  \beta^2-4)b+3 \alpha g}{12 b^2+g^2} \frac{\varphi_x^2+\varphi_y^2}{k_0},\
}\\ \displaystyle{ 
\frac{b}{g}=\frac{\beta^2-2}{6 \alpha-12 \beta}=\frac{2-\alpha^2}{6 \beta-12 \alpha},\
\varphi_x^2+\varphi_y^2 \not=0,
}
\end{array}
\right.
\label{eqSBS444tau0dominant} 
\end{eqnarray}
and the two phases $a_1,a_2$ are arbitrary functions.
This implies the two mutually exclusive possibilities
\begin{eqnarray}
& &
(\alpha+\beta=0) \hbox{ or } (\alpha^2-3 \alpha \beta+\beta^2+2=0).
\end{eqnarray}
The first one contains the unphysical reduction $U_2=\barU_1$,
while the second one describes a truly coupled behaviour.
In both cases, just like in \cite{CM2000b},
the eight Fuchs indices are $1,0,0$ (respectively corresponding to the 
arbitrary functions $\varphi_0$, $a_1$, $a_2$)
and five irrational values,
therefore there is no no-log condition to compute.

\subsection{Case $g=0$}
\label{sectionPtestg0}

For $g=0$, 
the system (\ref{eqSBS444}) is made of two coupled nonlinear Schr\"odinger equations
in 2+1 dimensions,
of a type which is nonintegrable \cite{FK1983}
but for which some closed form solutions have been found \cite{Khukhunashvili}. 
The system admits the same simple pole behaviour as (\ref{eqSBS466gtau}), for $U_1,\barU_1,U_2,\barU_2$ 
\begin{eqnarray}
& & {\hskip -10.0truemm}
    U_1 \sim M e^{ i a_1} \chi^{-1},\
\barU_1 \sim M e^{-i a_1} \chi^{-1},\
    U_2 \sim M e^{ i a_2} \chi^{-1},\
\barU_2 \sim M e^{-i a_2} \chi^{-1},\
M^2=-\frac{\varphi_x^2+\varphi_y^2}{3 k_0 b}\cdot
\label{eqSBS444g0} 
\end{eqnarray}
The eight Fuchs indices $j$ are then the roots of $j(3-j)=-4,0,0,4/3$, i.e.
\begin{eqnarray}
& &
-1,0,0,3,3,4,(3+\sqrt{3})/2,(3-\sqrt{3})/2,
\end{eqnarray}
and the presence of noninteger indices is sufficient to prove the nonintegrability.

After a computation quite similar to that of section \ref{sectionPtestgeneric},
we also could not find at least one particular solution to the set of five no-log conditions.

To conclude this local analysis (of the movable singularities),
closed form singlevalued solutions are not impossible to find,
but this will prove quite difficult.
Such an investigation is performed in section \ref{sectionOne-family}.

 
\section{Search for radial shock-type solutions, generic case $g \tau \not=0$}
\label{sectionOne-family}

For the radial reduction $\partial_\theta=0$ suggested by the local analysis,
see (\ref{eqSBS466gtau_Q1aQ1breducsol}),
let us look for possible closed form singlevalued solutions defined by the assumption
\begin{eqnarray}
& &
\bfu =\sum_{j=0}^{1} \bfu_j \chi^{j-1},\
\chi=\varphi-\varphi_0,
\label{eqOne-family.u} 
\end{eqnarray}
i.e.
\begin{eqnarray}
& &
\left\lbrace
\begin{array}{ll}
\displaystyle{
    U_1 = M e^{ i a_1} (\chi^{-1}+    U_{1,1}),\
\barU_1 = M e^{-i a_1} (\chi^{-1}+\barU_{1,1}),\
}\\ \displaystyle{
    U_2 = M e^{ i a_2} (\chi^{-1}+    U_{2,1}),\
\barU_2 = M e^{-i a_2} (\chi^{-1}+\barU_{2,1}),\
}\\ \displaystyle{
    Q   = N e^{i a_1-i a_2} (\chi^{-1}+    Q_{1}),\
\barQ   = N e^{i a_2-i a_1} (\chi^{-1}+\barQ_{1}),\
}
\end{array}
\right.
\label{eqSBS466gtauOne-family-truncation} 
\end{eqnarray}
in which the functions $M,N,a_1,a_2,\varphi$ must obey the relations
(\ref{eqSBS466gtau}) and (\ref{eqSBS466gtau_nolog}),
and the functions $U_{1,1},U_{2,1},Q_1$ are to be determined.
When one inserts such an assumption into the 
six equations of the system (\ref{eqSBS466}),
which we denote $\bfE=0$,
one generates a Laurent series which also terminates
\begin{eqnarray} 
& & {\hskip -14.0truemm}
\bfE=\sum_{j=0}^{-\bfq} \bfE_j \chi^{j+\bfq},\ \bfq=(-3,-3,-3,-3,-2,-2),
\end{eqnarray}
and the method is to solve the set of 11 complex equations
\begin{eqnarray} 
& & {\hskip -14.0truemm}
\forall j:\ \bfE_j=0
\end{eqnarray} 
for the 11 unknowns $M,N,a_1,a_2,\varphi$ (real) and $U_{1,1},U_{2,1},Q_1$ (complex).
This is the famous ``one-family truncation'' initiated by Weiss \textit{et al.} \cite{WTC},
see details in \cite{CMBook}. The result is the following.

The three complex equations $j=0$ first provide $M,N$ as in (\ref{eqSBS466gtau}),
and $a_1,a_2$ remain arbitrary (because $j=0$ is a double Fuchs index).

The next three complex equations $j=1$ yield the two conditions $Q_{1,a}=0$, $Q_{1,b}=0$, 
see (\ref{eqSBS466gtau_Q1b}),
a unique value for $U_{1,1},U_{2,1}$,
and an arbitrary value for $Q_1$.

At $j=2$, one obtains a unique value for $Q_1$,
plus four real constraints on $\varphi,a_1,a_2$.

Finally, the two complex equations $j=3$ yields four more real such constraints, 
among them the two no-log conditions $Q_{3,a}=0$, $Q_{3,b}=0$.

If one now assumes the three complex amplitudes to be independent 
of the polar coordinate $\theta$,
in order to proceed one has to choose (\ref{eqSBS466gtau_Q1aQ1breducsol}).
Then one of the four constraints $j=2$ 
yields
\begin{eqnarray} 
& & {\hskip -14.0truemm}
a_1+a_2=R g_1(z,t) + g_0(z,t),
\end{eqnarray}
in which $g_1$ and $g_0$ are arbitrary functions of two variables.

Finally, one of the three remaining constraints at $j=2$ yields the condition
\begin{eqnarray} 
& & {\hskip -14.0truemm}
 g_1(z,t)^2=\hbox{rational}(z;G_0(\varphi,z),G_1(\varphi,z)),
\nonumber 
\end{eqnarray} 
in which the rhs is a rational function of $z$ with coefficients depending on $G_0,G_1$,
and the condition that the rhs be independent of $\varphi$
admits no solution. 
Therefore a solution described by one family such as (\ref{eqSBS466gtauOne-family-truncation})
probably does not exist when one waives the restriction $\partial_\theta=0$.

Because of these difficulties, we did not try to find possible
pulse solutions described by the two-family truncation \cite[\S 5.7.2]{CMBook}.

\section{Lie symmetries}
\label{sectionClassicalSymmetries}

For convenience, we denote the independent variales $x,y,z,t$ as $x_j,j=1,2,3,4$
and the dependent variables $U_1,\barU_1,U_2,\barU_2,Q,\barQ$ as $u_k,k=1,\cdots,6$.

The method of Lie consists in unveiling the invariance properties of a given 
system of PDEs,
in order to define reductions to another system with a lesser number of independent
variables.
We refer the reader to pedagogical textbooks such as
\cite{OlverBook}
\cite{OvsiannikovBook} 
\cite{IbragimovBook},
and to a recent paper \cite{MMO2013} handling an example in full detail.

In order to apply the classical method to the system (\ref{eqSBS466}), 
we consider the one-parameter Lie group of infinitesimal transformations
\begin{eqnarray}
& &
\left\lbrace
\begin{array}{ll}
\displaystyle{
x_j^*=x_j+\varepsilon  \xi_j(x_1,x_2,x_3,x_4,u_1,u_2,u_3,u_4,u_5,u_6),\
}\\ \displaystyle{
u_k^*=u_k+\varepsilon \eta_k(x_1,x_2,x_3,x_4,u_1,u_2,u_3,u_4,u_5,u_6),\
}
\end{array}
\right.
\label{eq466Infinitesimal} 
\end{eqnarray}
where $\varepsilon$ is the group parameter.

One then requires this transformation to leave invariant the set
of solutions of the system (\ref{eqSBS466}). 
This yields an overdetermined, linear system of equations 
(called determining equations) for the infinitesimals $\xi_j,\eta_k$.
Having determined the infinitesimals, 
the symmetry variables are found by solving the invariant surface condition
\begin{equation}
\label{sur}
\Phi \equiv \sum_{j=1}^{4} \xi_j  \frac{\partial f}{\partial x_j}
          + \sum_{k=1}^{6} \eta_k \frac{\partial f}{\partial u_k}
          - \varphi=0.
\end{equation}

The associated Lie algebra of infinitesimal symmetries is the set
of vector fields of the form
\begin{equation}
\label{vect}
\begin{array}{l} 
\bfv=
  \xi_1 \partial_{x_1}+ \xi_2 \partial_{x_2}+ \xi_3 \partial_{x_3}+ \xi_4 \partial_{x_4}
+\eta_1 \partial_{u_1}+\eta_2 \partial_{u_2}+\eta_3 \partial_{u_3}
+\eta_4 \partial_{u_4}+\eta_5 \partial_{u_5}+\eta_6 \partial_{u_6}.
\end{array}
\end{equation}


In the generic case $g \tau \vg \not=0$ (4 independent variables, 6 real dependent variables, 6 real equations),
system (\ref{eqSBS466}) leads to a set of 65 determining equations, 
whose solution defines a Lie algebra with 10 generators,
\begin{eqnarray}
& & {\hskip -15.0truemm}
\left\lbrace
\begin{array}{ll}
\displaystyle{
T_x=\partial_x,\
T_y=\partial_y,\
T_z=\partial_z,\
T_t=\partial_t,\
\Theta= x \partial_y -y \partial_x,\
}\\ \displaystyle{
\genEx=z\partial_x+i k_0 x B_6,\ 
}\\ \displaystyle{
\genEy=z\partial_y+i k_0 y B_6,\ 
}\\ \displaystyle{         
B_6=                    u_1\partial_{u_1}-u_2\partial_{u_2}-u_3\partial_{u_3}+u_4\partial_{u_4}+2 u_5\partial_{u_5}-2 u_6\partial_{u_6},\
}\\ \displaystyle{
B_4=                    u_1\partial_{u_1}-u_2\partial_{u_2}+u_3\partial_{u_3}-u_4\partial_{u_4},\
}\\ \displaystyle{
\genA   = \vg z B_6 -t B_4. 
}
\end{array}
\right.
\label{eq466Lie_algebra} 
\end{eqnarray}

In the nongeneric case $g \tau\not=0, \vg =0$, one obtains the 11 generators
\begin{eqnarray}
& &
\left\lbrace
\begin{array}{ll}
\displaystyle{
T_x,\ T_y,\ T_z,\ T_t,\ \Theta,\ \genEx,\ \genEy,\ B_6,\ F(t) B_4,\
}\\ \displaystyle{
\gVsix=x \partial_x +y \partial_y +2 z \partial_z 
          -(u_1\partial_{u_1}+u_2\partial_{u_2}+u_3\partial_{u_3}+u_4\partial_{u_4}+2 u_5\partial_{u_5}+2 u_6\partial_{u_6}),\
}\\ \displaystyle{
\gWsix=z \gVsix -z^2 \partial_z+\frac{i k_0}{2} (x^2+y^2) B_6,\
}
\end{array}
\right.
\label{eq466vg0Lie_algebra} 
\end{eqnarray}
in which $F$ is an arbitrary function of one variable. 

The three other nongeneric cases $g \tau=0$ (i.e.~$g=0$ or $\tau=0$ or $g=\tau=0$)
admit the same set of generators independent of $g$ and $\tau$,
and these depend on whether $\vg\not=0$ or $\vg=0$.
For $g \tau=0, \vg\not=0$, introducing the notation
\begin{eqnarray}
& &
\left\lbrace
\begin{array}{ll}
\displaystyle{
\gBone=u_1\partial_{u_1}-u_2 \partial_{u_2},\ 
}\\ \displaystyle{
\gBtwo=u_3\partial_{u_3}-u_4 \partial_{u_4},\ 
}
\end{array}
\right.
\label{eqx44Notation} 
\end{eqnarray}
there exist 14 generators 
\begin{eqnarray}
& &
\left\lbrace
\begin{array}{ll}
\displaystyle{
T_x,\ T_y,\ T_z,\ T_t,\ \Theta,\ F_1(t- \vg z)\gBone,\ F_2(t+ \vg z) \gBtwo,\
}\\ \displaystyle{
\genex=z \partial_x+ i k_0 x (\gBone - \gBtwo),\
}\\ \displaystyle{
\geney=z \partial_y+ i k_0 y (\gBone - \gBtwo),\
}\\ \displaystyle{
\gVfou=x \partial_x+y \partial_y+2 z \partial_z+2 t \partial_t
 -(u_1\partial_{u_1}+u_2\partial_{u_2}+u_3\partial_{u_3}+u_4\partial_{u_4}),\
}\\ \displaystyle{
\genGx= t \partial_x + i k_0 \vg x (\gBone + \gBtwo),\
}\\ \displaystyle{
\genGy= t \partial_y + i k_0 \vg y (\gBone + \gBtwo),\
}\\ \displaystyle{
\genHx=\frac{1}{2}(\vg^2 z^2-t^2) \partial_x + i k_0 \vg x [\vg z (\gBone-\gBtwo) -t (\gBone+\gBtwo)],\ 
}\\ \displaystyle{
\genHy=\frac{1}{2}(\vg^2 z^2-t^2) \partial_y + i k_0 \vg y [\vg z (\gBone-\gBtwo) -t (\gBone+\gBtwo)],\ 
}
\end{array}
\right.
\label{eq444Lie_algebra} 
\end{eqnarray}
in which $F_1,F_2$ are arbitrary functions, 
while
for $g \tau=0, \vg=0$, there exist only 10 generators
\begin{eqnarray}
& &
\left\lbrace
\begin{array}{ll}
\displaystyle{
T_x,\ T_y,\ T_z,\ \Theta,\ \gBone,\ \gBtwo,\ \genex,\ \geney,\ 
}\\ \displaystyle{
\gVthr=x \partial_x+y \partial_y+2 z \partial_z
 -(u_1\partial_{u_1}+u_2\partial_{u_2}+u_3\partial_{u_3}+u_4\partial_{u_4}),\
}\\ \displaystyle{
\gWfou= z \gVthr -z^2 \partial_z +\frac{i k_0}{2}(x^2+y^2)(\gBone - \gBtwo). 
}
\end{array}
\right.
\label{eq344Lie_algebra} 
\end{eqnarray}

Let us first define the shorthand notation
\begin{eqnarray}
& &
E_{n}^\pm= i k_0 \left[(t-\vg z)^n F_1(t-\vg z)\gBone\pm(t+\vg z)^n F_2(t+\vg z)\gBtwo\right],\ n=0,1,2.
\label{eqNotationEnpm}
\end{eqnarray}

For each of the four algebras 
(\ref{eq466Lie_algebra}), (\ref{eq466vg0Lie_algebra}), (\ref{eq444Lie_algebra}), (\ref{eq344Lie_algebra}), 
we have built the commutator tables 
Table~\ref{TableCommutator466}
(gathering both (\ref{eq466Lie_algebra}) and (\ref{eq466vg0Lie_algebra})),
Table~\ref{TableCommutator444},
Table~\ref{TableCommutator344},
and the adjoint tables 
Table~\ref{TableAdjoint466}
(gathering both (\ref{eq466Lie_algebra}) and (\ref{eq466vg0Lie_algebra})),
Table~\ref{TableAdjoint444},
Table~\ref{TableAdjoint344},
which show the separate adjoint actions of
each element in a Lie algebra, as it acts on all other elements. 
This construction is done by summing the Lie series 
with the Baker-Campbell-Hausdorf formula 
\begin{eqnarray}
& &
e^{-\varepsilon X} Y e^{\varepsilon X} = Y
-      \varepsilon              [X,Y]
+\frac{\varepsilon^2}{2!}    [X,[X,Y]]
-\frac{\varepsilon^3}{3!} [X,[X,[X,Y]]] + \cdots
\end{eqnarray}


\section{Reductions}
\label{sectionClassicalReductions}

Let us give a few examples of such reductions.





\subsection{Reductions, generic case $g \tau \vg\not=0$ (10 generators)} 

The most general generator (the coefficients $a_k$ denote complex constants)
\begin{eqnarray}
& &
a_x T_x + a_y T_y + a_z T_z + a_t T_t +a_1 \Theta + a_2 \genEx + a_3 \genEy +a_4 B_4 + a_6 B_6 + a_0 A
\nonumber\\ & &
=\left[a_2 z - a_1 y + a_x\right] \partial_x   
+\left[a_3 z + a_1 x + a_y\right] \partial_y   
+      a_z                        \partial_z   
+      a_t                        \partial_t  
\nonumber\\ & & 
+E_1 ( u_1\partial_{u_1}  -u_2\partial_{u_2})
+E_2 (-u_3\partial_{u_3}  +u_4\partial_{u_4})
+E_3(2 u_5\partial_{u_5}-2 u_6\partial_{u_6}),
\label{eq466Generator}
\end{eqnarray}
in which
\begin{eqnarray}
& &
E_1= i k_0 (a_2 x + a_3 y) + a_6 + a_0 (\vg z -t) + a_4,
\nonumber\\ & &
E_2= i k_0 (a_2 x + a_3 y) + a_6 + a_0 (\vg z +t) - a_4,
\nonumber\\ & & 
E_3= i k_0 (a_2 x + a_3 y) + a_6 + a_0  \vg z,          
\end{eqnarray}
defines the first order characteristic system
\begin{eqnarray}
& &
  \frac{\D x}{a_2 z -a_1 y + a_x}
= \frac{\D y}{a_3 z +a_1 x + a_y}
= \frac{\D z}{a_z}
= \frac{\D t}{a_t}
\nonumber\\ & &
= \frac{\D u_1}{E_1 u_1}=-\frac{\D u_2}{E_1 u_2}
=-\frac{\D u_3}{E_2 u_3}= \frac{\D u_4}{E_2 u_4}
= \frac{\D u_5}{2 E_3 u_5}=-\frac{\D u_6}{2 E_3 u_6}.
\label{eq466CharSys}
\end{eqnarray}
The three equations in the first line of (\ref{eq466CharSys}) can be integrated 
and define three invariants only depending on $x,y,z,t$.
When $a_1 a_z$ is nonzero, these are 
[$(\xi_1,\xi_2)$ are chosen so evaluate to $(x,y)$ 
when $a_2=a_3=a_x=a_y=0$ and $z=0$],
\begin{eqnarray}
& & {\hskip -15.0 truemm}
a_1 a_z\not=0:
\left\lbrace
\begin{array}{ll}
\displaystyle{
\xi_1=\frac{1}{a_1^2}\left\lbrace
+[a_1 (a_3 z + a_1 x + a_y) -a_2 a_z] \cos\frac{a_1 z}{a_z}
-[a_1 (a_2 z - a_1 y + a_x) +a_3 a_z] \sin\frac{a_1 z}{a_z}\right\rbrace,   
}\\ \displaystyle{
\xi_2=\frac{1}{a_1^2}\left\lbrace
-[a_1 (a_3 z + a_1 x + a_y) -a_2 a_z] \sin\frac{a_1 z}{a_z}
-[a_1 (a_2 z - a_1 y + a_x) +a_3 a_z] \cos\frac{a_1 z}{a_z}\right\rbrace,
}\\ \displaystyle{
\xi_3=a_z t -a_t z.
}
\end{array}
\right.
\label{eqSBS466red0}
\end{eqnarray}
When one sets $a_0=a_4=0$, six other invariants can be found.
Indeed, the three expressions $E_1,E_2,E_3$ are then equal and the characteristic system
(\ref{eq466CharSys}) implies 
\begin{eqnarray}
& &
a_1 \frac{\D u_1}{u_1}-i k_0 \D (-a_3 x + a_2 y)
=\left[a_1 a_6 -i k_0 (-a_3 a_x+a_2 a_y)\right] \frac{\D z}{a_z}.
\end{eqnarray}
The corresponding reduction
\begin{eqnarray}
& & a_1 a_z\not=0, a_0=a_4=0:
\left\lbrace
\begin{array}{ll}
\displaystyle{
    U_1=    V_1(\xi_1,\xi_2,\xi_3)\ e^{   i k_0 F},\
    U_2=    V_2(\xi_1,\xi_2,\xi_3)\ e^{-  i k_0 F},\
}\\ \displaystyle{
    Q  =    W  (\xi_1,\xi_2,\xi_3)\ e^{ 2 i k_0 F},\
    F=\frac{-a_3 x + a_2 y}{a_1}-\frac{a_2^2+a_3^3}{2 a_1^2} z,
}
\end{array}
\right.
\end{eqnarray}
yields the reduced system, 
\begin{eqnarray}
& & a_1 a_z\not=0:
\left\lbrace
\begin{array}{ll}
\displaystyle{\phantom{-}
   i (-a_t+a_z \vg) V_{1,\xi_3}+\frac{V_{1,\xi_1 \xi_1}+V_{1,\xi_2 \xi_2}}{2 k_0}
 +\frac{a_1}{a_z} (\xi_2 V_{1,\xi_1}-\xi_1 V_{1,\xi_2})
}\\ \displaystyle{\phantom{1234}
 + b \left(\mod{V_1}^2+2 \mod{V_2}^2 \right) V_1 + i \frac{g}{2} W V_2=0,
}\\ \displaystyle{
 -i (-a_t-a_z \vg) V_{2,\xi_3}+\frac{V_{2,\xi_1 \xi_1}+V_{2,\xi_2 \xi_2}}{2 k_0} 
 +\frac{a_1}{a_z} (\xi_2 V_{2,\xi_1}-\xi_1 V_{2,\xi_2})
}\\ \displaystyle{\phantom{1234}
 + b \left(\mod{V_2}^2+2 \mod{V_1}^2 \right) V_2 - i \frac{g}{2}\barW V_1=0,
}\\ \displaystyle{
 a_z \tau W_{\xi_3} +W - V_1 \barV_2=0.
}
\end{array}
\right.
\end{eqnarray}

Another reduction is obtained by choosing the constants in (\ref{eq466CharSys})
as follows,
\begin{eqnarray}
& &
\frac{a_z}{a_1}=\frac{a_x}{a_3}=\frac{a_y}{a_2},\ a_x^2=a_y^2.
\end{eqnarray}
This reduction is defined by
\begin{eqnarray}
& &
\left\lbrace
\begin{array}{ll}
\displaystyle{

    U_1=    V_1(\xi,z,t)\ e^{   i k_0 F},\
    U_2=    V_2(\xi,z,t)\ e^{-  i k_0 F},\
    Q  =    W  (\xi,z,t)\ e^{ 2 i k_0 F},\
}\\ \displaystyle{
    \xi=\frac{a_1}{2}(x^2+y^2) + (a_3 x -a_2 y) z,\
    F=\frac{-a_3 x + a_2 y}{a_1},
}
\end{array}
\right.
\end{eqnarray}
and the reduced system
\begin{eqnarray}
& &
\left\lbrace
\begin{array}{ll}
\displaystyle{
\phantom{+}
 i (V_{1,z} + \vg V_{1,t}) +(2 a_1 \xi + A^2 z^2)\frac{V_{1,\xi\xi}}{2 k_0}
 + (\frac{a_1^2}{k_0}- i k_0 \frac{A^2}{a_1} z) V_{1,\xi} - \frac{A^2 k_0}{2 a_1^2} V_1
 + b \left(\mod{V_1}^2+2 \mod{V_2}^2 \right) V_1 + i \frac{g}{2} W V_2=0,
}\\ \displaystyle{
-i (V_{2,z} - \vg V_{2,t}) +(2 a_1 \xi + A^2 z^2)\frac{V_{2,\xi\xi}}{2 k_0}
 + (\frac{a_1^2}{k_0}+ i k_0 \frac{A^2}{a_1} z) V_{2,\xi} - \frac{A^2 k_0}{2 a_1^2} V_2
  + b \left(\mod{V_2}^2+2 \mod{V_1}^2 \right) V_2 - i \frac{g}{2}\barW V_1=0,
}\\ \displaystyle{
 \tau W_t +W - V_1 \barV_2=0,
}
\end{array}
\right.
\label{eqSBS466red1} 
\end{eqnarray}
depends on one more arbitrary constant, $A^2=a_2^2+a_3^2$, since $a_1$ can be set to any nonzero numerical value.
When $A=0$ (i.e.~$a_2=a_3=0$),
this reduction is identical to the radial reduction $\partial_\theta=0$ in polar coordinates
\begin{eqnarray}
& &
x=\rho \cos \theta,\ y=\rho \sin \theta.
\end{eqnarray}

\subsection{Nongeneric case $\vg=0, g \tau \not=0$ (11 generators)} 

\begin{eqnarray}
& &
a_x T_x + a_y T_y + a_z T_z + a_t T_t +a_1 \Theta + a_2 \genEx + a_3 \genEy +a_6 B_6 
+b_4 F(t) B_4+ c_6 \gVsix + d_6 \gWsix
\nonumber\\ & & 
=\left[a_2 z - a_1 y + a_x +   c_6 x + d_6 z x\right] \partial_x   
+\left[a_3 z + a_1 x + a_y +   c_6 y + d_6 z y\right] \partial_y   
+\left[                a_z + 2 c_6 z + d_6 z^2\right] \partial_z   
+                      a_t                            \partial_t  
\nonumber\\ & & 
+E_1 (  u_1\partial_{u_1}-  u_2\partial_{u_2})
+E_2 (- u_3\partial_{u_3}+  u_4\partial_{u_4})
+E_3 (2 u_5\partial_{u_5}-2 u_6\partial_{u_6})
\nonumber\\ & & 
+(c_6+d_6 z) (u_1\partial_{u_1}+ u_2\partial_{u_2}+u_3\partial_{u_3}+ u_4\partial_{u_4}
+2 u_5\partial_{u_5}+2 u_6\partial_{u_6}),
\label{eq466vg0Generator}
\end{eqnarray}
in which
\begin{eqnarray}
& &
E_1= i k_0 (a_2 x + a_3 y) + a_6 + d_6 i k_0 (x^2+y^2)/2 + b_4 F(t) ,
\nonumber\\ & &
E_2= i k_0 (a_2 x + a_3 y) + a_6 + d_6 i k_0 (x^2+y^2)/2- b_4 F(t) ,
\nonumber\\ & & 
E_3= i k_0 (a_2 x + a_3 y) + a_6 + d_6 i k_0 (x^2+y^2)/2,         
\end{eqnarray}
defines the first order characteristic system
\begin{eqnarray}
& &
  \frac{\D x}{a_2 z -a_1 y + a_x+   c_6 x + d_6 z x}
= \frac{\D y}{a_3 z +a_1 x + a_y+   c_6 y + d_6 z y}
= \frac{\D z}{a_z               + 2 c_6 z + d_6 z^2}
= \frac{\D t}{a_t}
\nonumber\\ & &
= \frac{\D u_1}{( E_1+c_6+d_6 z) u_1}
= \frac{\D u_2}{(-E_1+c_6+d_6 z) u_2}
= \frac{\D u_3}{(-E_2+c_6+d_6 z) u_3}
= \frac{\D u_4}{( E_2+c_6+d_6 z) u_4}
\nonumber\\ & &
= \frac{\D u_5}{2 (E_3+c_6+d_6 z) u_5}
= \frac{\D u_6}{-2 (E_3+c_6+d_6 z) u_6}.
\label{eq466vg0CharSys}
\end{eqnarray}

If one requires the sought after reduction to be noncharacteristic
(i.e.~to preserve the total differential order ten),
it is quite difficult to find such a reduction.

\vfill\eject

\vspace{20pt}
\begin{table}[ht]
\caption{466. Commutator table for the Lie algebras (\ref{eq466Lie_algebra}) and (\ref{eq466vg0Lie_algebra}). 
Example: $[T_x,\Theta]=T_y$.
The 10-dim Lie algebra (\ref{eq466Lie_algebra}) is recovered
by suppressing the two lines and columns labelled $\gVsix$ and $\gWsix$
and setting $F(t)=1$.
For the 11-dim Lie algebra (\ref{eq466vg0Lie_algebra}),
suppress the line and column labelled $\genA$.
\label{TableCommutator466}
}
\begin{center}
\footnotesize
\begin{tabular}{|l|ll|llllllllll|}
\hline &$\genA$ &$\Theta$ & $T_z$  & $T_t$ & $T_x$ & $T_y$ &$\genEx$ &$\genEy$ &$B_6$ &$F(t)B_4$ &$\gVsix$ & $\gWsix$
\\[10pt]\hline
$\genA$ &$0$ &$0$ &$-\vg B_6$ &$B_4$ &$0$ &$0$ &$0$ &$0$ &$0$ &$0$ &N/A &N/A 
\\[10pt]
$\Theta$ &$0$ &$0$ &$0$ &$0$ &$-T_y$ &$T_x$ &$-\genEy$ &$\genEx$ &$0$ &$0$ &$0$ &$0$ 
\\[10pt]\hline
$T_z$ &$\vg B_6$ &$0$ &$0$ &$0$ &$0$ &$0$ &$T_x$ &$T_y$ &$0$ &$$ &$2 T_z$ &$\gVsix$
\\[10pt]
$T_t$ &$-B_4$ &$0$ &$0$ &$0$ &$0$ &$0$ &$0$ &$0$ &$0$ &$F'(t) B_4$ &$0$ &$0$ 
\\[5pt]
$T_x$ &$0$ &$ T_y$ &$0$ &$0$ &$0$ &$0$ &$i k_0 B_6$ &$0$ &$0$ &$0$ &$T_x$ &$\genEx$
\\[10pt]
$T_y$ &$0$ &$-T_x$ &$0$ &$0$ &$0$ &$0$ &$0$ &$i k_0 B_6$ &$0$ &$0$ &$T_y$ &$\genEy$ 
\\[10pt]
$\genEx$ &$0$ &$ \genEy$ &$-T_x$ &$0$ &$-i k_0 B_6$ &$0$ &$0$ &$0$ &$0$ &$0$ &$-\genEx$ &$0$ 
\\[10pt]
$\genEy$ &$0$ &$-\genEx$ &$-T_y$ &$0$ &$0$ &$-i k_0 B_6$ &$0$ &$0$ &$0$ &$0$ &$-\genEy$ &$0$ 
\\[5pt]
$B_6$ &$0$ &$0$ &$0$ &$0$ &$0$ &$0$ &$0$ &$0$ &$0$ &$0$ &$0$ &$0$ 
\\[5pt]
$F(t)B_4$ &$0$ &$0$ &$0$ &$-F'(t) B_4$ &$0$ &$0$ &$0$ &$0$ &$0$ &$0$ &$0$ &$0$ 
\\[5pt]
$\gVsix$ &N/A &$0$ &$-2 T_z$ &$0$ &$-T_x$ &$-T_y$ &$\genEx$ &$\genEy$ &$0$ &$0$ &$0$ &$2 \gWsix$ 
\\[5pt]
$\gWsix $&N/A &$0$ &$-\gVsix$ &$0$ &$-\genEx$ &$-\genEy$ &$0$ &$0$ &$0$ &$0$ &$-2 \gWsix$ &$0$ 
\\[5pt]\hline
\end{tabular}
\end{center}
\end{table}


\vspace{20pt}
\begin{landscape}
\begin{table}[ht]
\caption{444. Commutator table for the Lie algebra (\ref{eq444Lie_algebra}). 
The abbreviation $E_{n}^\pm$ is defined in (\ref{eqNotationEnpm}).
\label{TableCommutator444}
}
\begin{center}
\footnotesize
\begin{tabular}{|l|llllllllllllll|}
\hline &$T_x$ &$T_y$ &$T_z$ &$T_t$ &$\Theta$ &$F_1\gBone$ &$F_2\gBtwo$ &$\genex$ &$\geney$ &$\gVfou$ &$\genGx$ &$\genGy$ &$\genHx$ &$\genHy$
\\[5pt]\hline
$T_x$ &$0$ &$0$ &$0$ &$0$ &$T_y$ &$0$ &$0$ &$i k_0(\gBone-\gBtwo)$ &$0$ &$T_x$ &$i k_0\vg(\gBone+\gBtwo)$ &$0$ &$[T_x,\genHx]$ &$0$
\\[5pt]
$T_y$ &$0$ &$0$ &$0$ &$0$ &$-T_x$ &$0$ &$0$ &$0$ &$[T_x,\genex]$ &$T_y$ &$0$ &$[T_x,\genGx]$ &$0$ &$[T_x,\genHx]$
\\[5pt]
$T_z$ &$0$ &$0$ &$0$ &$0$ &$0$ &$-\vg\gBone$ &$-\vg\gBtwo$ &$T_x$ &$T_y$ &$2T_z$ &$0$ &$0$ &$\vg^2\genex$ &$\vg^2\geney$
\\[5pt]
$T_t$ &$0$ &$0$ &$0$ &$0$ &$0$ &$F_1'\gBone$ &$F_2'\gBtwo$ &$0$ &$0$ &$2T_t$ &$T_x$ &$T_y$ &$-\genGx$ &$-\genGy$
\\[5pt]
$\Theta$ &$-T_y$ &$T_x$ &$0$ &$0$ &$0$ &$F_1\gBone$ &$F_2\gBtwo$ &$-\geney$ &$\genex$ &$0$ &$-\genGy$ &$\genGx$ &$-\genHy$ &$\genHx$
\\[5pt]
$F_1\gBone$ &$0$ &$0$ &$\vg\gBone$ &$-F_1'\gBone$ &$-F_1\gBone$ &$0$ &$0$ &$0$ &$0$ &$-2F_1'\gBone$ &$0$ &$0$ &$0$ &$0$
\\[5pt]
$F_2\gBtwo$ &$0$ &$0$ &$\vg\gBtwo$ &$-F_2'\gBtwo$ &$-F_2\gBtwo$ &$0$ &$0$ &$0$ &$0$ &$-2F_2'\gBtwo$ &$0$ &$0$ &$0$ &$0$
\\[5pt]
$\genex$ &$-i k_0(\gBone-\gBtwo)$ &$0$ &$-T_x$ &$0$ &$\geney$ &$0$ &$0$ &$0$ &$0$ &$-\genex$ &$[\genex,\genGx]$ &$0$ &$[\genex,\genHx]$ &$0$
\\[5pt]
$\geney$ &$0$ &$-[T_x,\genex]$ &$-T_y$ &$0$ &$-\genex$ &$0$ &$0$ &$0$ &$0$ &$-\geney$ &$0$ &$[\genex,\genGx]$ &$0$ &$[\genex,\genHx]$
\\[5pt]
$\gVfou$ &$-T_x$ &$-T_y$ &$-2T_z$ &$-2T_t$ &$0$ &$2F_1'\gBone$ &$2F_2'\gBtwo$ &$\genex$ &$\geney$ &$0$ &$\genGx$ &$\genGy$ &$3\genHx$ &$3\genHy$
\\[5pt]
$\genGx$ &$-i k_0\vg(\gBone+\gBtwo)$ &$0$ &$0$ &$-T_x$ &$\genGy$ &$0$ &$0$ &$-[\genex,\genGx]$ &$0$ &$-\genGx$ &$0$ &$0$ &$[\genGx,\genHx]$ &$0$
\\[5pt]
$\genGy$ &$0$ &$-[T_x,\genGx]$ &$0$ &$-T_y$ &$-\genGx$ &$0$ &$0$ &$0$ &$-[\genex,\genGx]$ &$-\genGy$ &$0$ &$0$ &$0$ &$[\genGx,\genHx]$
\\[5pt]
$\genHx$ &$-[T_x,\genHx]$ &$0$ &$-\vg^2\genex$ &$\genGx$ &$\genHy$ &$0$&$0$ &$-[\genex,\genHx]$ &$0$ &$-3\genHx$ &$-[\genGx,\genHx]$ &$0$ &$0$ &$0$
\\[5pt]
$\genHy$ &$0$ &$-[T_x,\genHx]$ &$-\vg^2\geney$ &$\genGy$ &$-\genHx$ &$0$ &$0$ &$0$ &$-[\genex,\genHx]$ &$-3\genHy$ &$0$ &$-[\genGx,\genHx]$ &$0$ &$0$
\\[5pt]\hline
\end{tabular}
\end{center}
\end{table}

\begin{eqnarray}
&&
\begin{array}{ll}
    \displaystyle{[\genex,\genGx]=-    E_{1}^{-},\    
}\\ \displaystyle{[T_x   ,\genHx]=-\vg E_{1}^{+},\    
}\\ \displaystyle{[\genex,\genHx]=     E_{2}^{-}/2,\  
}\\ \displaystyle{[\genGx,\genHx]=-\vg E_{2}^{+}/2,\  
}\\ \displaystyle{[T_x   ,\genGx]= \vg E_{0}^{+},\    
}\\ \displaystyle{[T_x   ,\genex]=     E_{0}^{-},\    
}
\end{array}
\label{TableCommutator444Overflow}
\end{eqnarray}
\end{landscape}

\vspace{20pt}
\begin{table}[ht]
\caption{344. Commutator table for the Lie algebra (\ref{eq344Lie_algebra}). 
\label{TableCommutator344}}
\begin{center}
\footnotesize
\begin{tabular}{|l|llllllllll|}
\hline &$T_x$ &$T_y$ &$T_z$ &$\Theta$ &$\gBone$ &$\gBtwo$ &$\genex$ &$\geney$ &$\gVthr$ &$\gWfou$
\\[5pt]\hline
$T_x$ &$0$ &$0$ &$0$ &$T_y$ &$0$ &$0$ &$i k_0(\gBone-\gBtwo)$ &$0$ &$T_x$ &$\genex$ 
\\[5pt]
$T_y$ &$0$ &$0$ &$0$ &$-T_x$ &$0$ &$0$ &$0$ &$i k_0(\gBone-\gBtwo)$ &$T_y$ &$\geney$
\\[5pt]
$T_z$ &$0$ &$0$ &$0$ &$0$ &$0$ &$0$ &$T_x$ &$T_y$ &$2T_z$ &$\gVthr$
\\[5pt]
$\Theta$ &$-T_y$ &$T_x$ &$0$ &$0$ &$0$ &$0$ &$-\geney$ &$\genex$ &$0$ &$0$
\\[5pt]
$\gBone$ &$0$ &$0$ &$0$ &$0$ &$0$ &$0$ &$0$ &$0$ &$0$ &$0$ 
\\[5pt]
$\gBtwo$ &$0$ &$0$ &$0$ &$0$ &$0$ &$0$ &$0$ &$0$ &$0$ &$0$ 
\\[5pt]
$\genex$ &$-i k_0(\gBone-\gBtwo)$ &$0$ &$-T_x$ &$\geney$ &$0$ &$0$ &$0$ &$0$ &$-\genex$ &$0$ 
\\[5pt]
$\geney$ &$0$ &$-i k_0(\gBone-\gBtwo)$ &$-T_y$ &$-\genex$ &$0$ &$0$ &$0$ &$0$ &$-\geney$ &$0$ 
\\[5pt]
$\gVthr$ &$-T_x$ &$-T_y$ &$-2T_z$ &$0$ &$0$ &$0$ &$\genex$ &$\geney$ &$0$ &$2\gWfou$
\\[5pt]
$\gWfou$ &$-\genex$ &$-\geney$ &$-\gVthr$ &$0$ &$0$ &$0$ &$0$ &$0$ &$-2\gWfou$ &$0$
\\[5pt]\hline
\end{tabular}
\end{center}
\end{table}

\vfill \eject
\vspace{20pt}
\begin{landscape}
\begin{table}[t]
\caption{466. Adjoint table for the Lie algebras (\ref{eq466Lie_algebra}) and (\ref{eq466vg0Lie_algebra}),
with the same convention than in Table \ref{TableCommutator466}.
The entry $(i,j)$ represents ${\rm Ad}(e^{\varepsilon v_i})|v_j$.
When it evaluates to $v_j$, it is simply represented by a dot (.) symbol.
The abbreviations $c$ and $s$ stand for $\cos\varepsilon$ and $\sin\varepsilon$.
Large expressions $(i,j)$ are listed just after the table.
N/A means not applicable.
\label{TableAdjoint466}
}
\begin{center}
\footnotesize
\begin{tabular}{|l|lll|lllllllll|}
\hline &$\genA$ &$\Theta$ &$T_z$  &$T_t$ &$T_x$ &$T_y$ &$\genEx$ &$\genEy$ &$B_6$ &$F(t)B_4$ &$\gVsix$ &$\gWsix$
\\[10pt]\hline
$\genA$ &$.$ &$.$ &$T_z+\varepsilon \vg B_6$ &$T_t-\varepsilon B_4$ &$.$ &$.$ &$.$ &$.$ &$.$ &$.$ &N/A &N/A 
\\[10pt]
$\Theta$ &$.$ &$.$ &$.$ &$.$ &$T_x c+T_y s$ &$T_y c-T_x s$ &$E_x c+E_y s$ &$E_y c- E_x s$ &$.$ &$.$ &$.$ &$.$ 
\\[10pt]
$T_z$    &$\genA-\varepsilon \vg B_6$ &$.$ &$.$ &$.$ &$.$ &$.$ &$E_x-\varepsilon T_x$ &$E_y-\varepsilon T_y$ &$.$ &$.$ &$\gVsix-2\varepsilon T_z$ &$(T_z,\gWsix)$
\\[10pt]\hline
$T_t$   &$(T_t,\genA)$ &$.$ &$.$ &$.$ &$.$ &$.$ &$.$ &$.$ &$.$ &$F B_4 e^{-\varepsilon F'/F}$  &$.$ &$.$
\\[5pt]
$T_x$   &$.$ &$\Theta-\varepsilon T_y$ &$.$ &$.$ &$.$ &$.$ &$E_x-\varepsilon i k_0 B_6$ &$.$ &$.$ &$.$ &$\gVsix-\varepsilon T_x$ &$(T_x,\gWsix)$
\\[10pt]
$T_y$    &$.$ &$\Theta+\varepsilon T_x$ &$.$ &$.$ &$.$ &$.$ &$.$ &$E_y-\varepsilon i k_0 B_6$ &$.$ &$.$ &$\gVsix-\varepsilon T_y$ &$(T_y,\gWsix)$
\\[10pt]
$\genEx$ &$.$ &$\Theta-\varepsilon E_y$ &$T_z+T_x \varepsilon+\varepsilon^2 i k_0 B_6/2$ &$.$ &$T_x+\varepsilon i k_0 B_6$ &$.$ &$.$ &$.$ &$.$ &$.$ &$\gVsix+\varepsilon E_x$ &$.$ 
\\[10pt]
$\genEy$ &$.$ &$\Theta+\varepsilon E_x$ &$T_z+T_y \varepsilon+\varepsilon^2 i k_0 B_6/2$ &$.$ &$.$ &$T_y+\varepsilon i k_0 B_6$ &$.$ &$.$ &$.$ &$.$ &$\gVsix+\varepsilon E_y$ &$.$ 
\\[5pt]
$B_6$      &$.$ &$.$ &$.$ &$.$ &$.$ &$.$ &$.$ &$.$ &$.$ &$.$ &$.$ &$.$
\\[5pt]
$F(t)B_4$  &$.$ &$.$ &$.$ &$T_t+\varepsilon B_4 F'$ &$.$ &$.$ &$.$ &$.$ &$.$ &$.$ &$.$ &$.$
\\[5pt]
$\gVsix$    &N/A &$.$ &$T_z e^{2\varepsilon}$ &$.$ &$T_x e^{\varepsilon}$ &$T_y e^{\varepsilon}$ &$E_x e^{-\varepsilon}$ &$E_y e^{-\varepsilon}$ &$.$ &$.$ &$.$ &$\gWsix e^{-2\varepsilon}$ 
\\[5pt]
$\gWsix$ &N/A &$.$ &$T_z+\varepsilon\gVsix+\varepsilon^2\gWsix$ &$.$ &$T_x+\varepsilon E_x$ &$T_y+\varepsilon E_y$ &$.$ &$.$ &$.$ &$.$ &$\gVsix+2\varepsilon \gWsix$ &$.$
\\[5pt]\hline
\end{tabular}
\end{center}
\end{table}

\begin{eqnarray}
&&
\begin{array}{ll}
\displaystyle{    (T_t,\genA)=\genA+B_4 (1-(F/F')e^{-\varepsilon F'/F}),
}\\ \displaystyle{(T_x,\gWsix)=\gWsix-\varepsilon E_x+\varepsilon^2 i k_0 B_6/2,
}\\ \displaystyle{(T_y,\gWsix)=\gWsix-\varepsilon E_y+\varepsilon^2 i k_0 B_6/2,
}\\ \displaystyle{(T_z,\gWsix)=\gWsix-\varepsilon \gVsix+ \varepsilon^2 T_z
}
\end{array}
\label{TableAdjointOverflow466}
\end{eqnarray}
   
\end{landscape}
\vfill\eject 

\tabcolsep=1.5truemm
\tabcolsep=0.5truemm
\tabcolsep=0.2truemm
 
\vspace{20pt}
\begin{landscape}
\begin{table}[t]
\caption{444. Adjoint table for the Lie algebra (\ref{eq444Lie_algebra}). 
The notation and convention are the same as in Table \ref{TableAdjoint466}.
The abbreviation $E_{n}^\pm$ is defined in (\ref{eqNotationEnpm}).
\label{TableAdjoint444}
}
\begin{center}
\footnotesize
\begin{tabular}{|l|llllllllllllll|}
\hline &$T_x$ &$T_y$ &$T_z$ &$T_t$ &$\Theta$ &$F_1\gBone$ &$F_2\gBtwo$ &$\genex$ &$\geney$ &$\gVfou$ &$\genGx$ &$\genGy$ &$\genHx$ &$\genHy$
\\[5pt]\hline
$T_x$ &$.$ &$.$ &$.$ &$.$ &$\Theta-\varepsilon T_y$ &$.$  &$.$ &$\genex-\varepsilon E_0^-$ &$.$ &$.$ 
      &$\genGx-\varepsilon \vg E_0^+$ &$.$ &$\genHx+\varepsilon E_1^+$ &$.$
\\[5pt]
$T_y$ &$.$ &$.$ &$.$ &$.$ &$\Theta+\varepsilon T_x$ &$.$  &$.$ &$.$ &$\geney-\varepsilon E_0^-$ &$.$ 
      &$.$ &$\genGy-\varepsilon \vg E_0^+$ &$.$ &$\genHy+\varepsilon E_1^+$
\\[5pt]
$T_z$ &$.$ &$.$ &$.$ &$.$ &$.$ &$F_1\gBone e^{\varepsilon \vg/F_1}$ &$F_2\gBtwo e^{\varepsilon \vg/F_2}$ 
      &$\genex-\varepsilon T_x$ &$\geney-\varepsilon T_y$ &$\gVfou-2\varepsilon T_z$ &$.$ &$.$ 
			&$(T_z,\genHx)$ &$(T_z,\genHy)$
\\[5pt]
$T_t$ &$.$ &$.$ &$.$ &$.$ &$.$ &$F_1\gBone e^{-\varepsilon F_1'/F_1}$ &$F_2\gBone e^{-\varepsilon F_2'/F_2}$ &$.$ &$.$ 
      &$\gVfou-2\varepsilon T_t$ &$\genGx-\varepsilon T_x$ &$\genGy-\varepsilon T_y$ 
			&$(T_t,\genHx)$ &$(T_t,\genHy)$
\\[5pt]
$\Theta$ &$T_x c+T_y s$ &$T_y c-T_x s$ &$.$ &$.$ &$.$ &$F_1\gBone e^{-\varepsilon}$ &$F_2\gBtwo e^{-\varepsilon}$ 
      &$\genex c+\geney s$ &$\geney c-\genex s$ &$.$ &$\genGx c+\genGy s$ &$\genGy c-\genGx s$ 
			&$\genHx c+\genHy s$ &$\genHy c-\genHx s$   
\\[5pt]
$F_1\gBone$ &$.$ &$.$ &$T_z-\varepsilon\vg\gBone$ &$T_t+\varepsilon F_1'\gBone$ &$\Theta+\varepsilon F_1\gBone$ &$.$ 
      &$.$ &$.$ &$.$ &$\gVfou+2\varepsilon F_1'\gBone$ &$.$ &$.$ &$.$ &$.$
\\[5pt]
$F_2\gBtwo$ &$.$ &$.$ &$T_z-\varepsilon\vg\gBtwo$ &$T_t+\varepsilon F_2'\gBtwo$ &$\Theta+\varepsilon F_2\gBtwo$ &$.$
      &$.$ &$.$ &$.$ &$\gVfou+2\varepsilon F_2'\gBtwo$ &$.$ &$.$ &$.$ &$.$
\\[5pt]
$\genex$    &$T_x+\varepsilon E_{0}^{-}$ &$.$ &$(\genex,T_z)$ &$.$ &$\Theta-\varepsilon\geney$ &$.$ 
      &$.$ &$.$ &$.$ &$\gVfou+\varepsilon\genex$ &$\genGx+\varepsilon E_{1}^{-}$ &$.$ &$\genHx-\varepsilon E_{2}^{-}/2$ &$.$
\\[5pt]
$\geney$    &$.$ &$T_y+\varepsilon E_{0}^{-}$ &$(\geney,T_z)$ &$.$ &$\Theta+\varepsilon\genex$ &$.$
      &$.$ &$.$ &$.$ &$\gVfou+\varepsilon\geney$ &$.$ &$\genGy+\varepsilon E_{1}^{-}$ &$.$ &$\genHy-\varepsilon E_{2}^{-}/2$
\\[5pt]
$\gVfou$     &$T_x e^\varepsilon$ &$T_y e^\varepsilon$ &$T_z e^{2\varepsilon}$ &$T_t e^{2\varepsilon}$ &$.$ 
  &$F_1\gBone e^{-2\varepsilon F_1'/F_1}$ &$F_2\gBtwo e^{-2\varepsilon F_2'/F_2}$ 
	&$\genex e^{-\varepsilon}$ &$\geney e^{-\varepsilon}$ &$.$ 
	&$\genGx e^{-\varepsilon}$ &$\genGy e^{-\varepsilon}$ &$\genHx e^{-3\varepsilon}$ &$\genHy e^{-3\varepsilon}$
\\[5pt]
$\genGx$    &$T_x+\varepsilon\vg E_{0}^{+}$ &$.$ &$.$ &$(\genGx,T_t)$ 
      &$\Theta-\varepsilon\genGy$ &$.$ &$.$ &$\genex-\varepsilon E_{1}^{-}$ &$.$ &$\gVfou+\varepsilon\genGx$ 
			&$.$ &$.$ &$\genHx+\varepsilon E_{2}^{+}/2$ &$.$
\\[5pt]
$\genGy$    &$.$ &$T_y+\varepsilon\vg E_{0}^{+}$ &$.$ &$(\genGy,T_t)$ 
      &$\Theta+\varepsilon\genGx$ &$.$ &$.$ &$.$ &$\geney-\varepsilon E_{1}^{-}$ &$\gVfou+\varepsilon\genGy$ 
			&$.$ &$.$ &$.$ &$\genHy-\varepsilon E_{2}^{+}/2$
\\[5pt]
$\genHx$    &$T_x-\varepsilon\vg E_{1}^{+}$ &$.$ &$(\genHx,T_z)$ 
      &$(\genHx,T_t)$ &$\Theta-\varepsilon\genHy$ &$.$ &$.$
			&$\genex+\varepsilon E_{2}^{-}/2$ &$.$ &$\gVfou+3\varepsilon\genHx$ &$\genGx-\varepsilon E_{2}^{+}/2$ &$.$ &$.$ &$.$
\\[5pt]
$\genHy$    &$.$ &$T_y-\varepsilon\vg E_{1}^{+}$ &$(\genHy,T_z)$ 
      &$(\genHy,T_t)$ &$\Theta+\varepsilon\genHx$ &$.$ &$.$ 
			&$.$ &$\geney+\varepsilon E_{2}^{-}/2$ &$\gVfou+3\varepsilon\genHy$ &$.$ &$\genGy+\varepsilon E_{2}^{+}/2$ &$.$ &$.$
\\[5pt]\hline
\end{tabular}
\end{center}
\end{table}

{\vglue -1.9truecm}

\begin{eqnarray}
&&
\begin{array}{ll}
\displaystyle{
                  (T_z,\genHx)=\genHx-\varepsilon \vg^2 \genex+\varepsilon^2 \vg^2 T_x/2,
}\\ \displaystyle{(T_z,\genHy)=\genHy-\varepsilon \vg^2 \geney+\varepsilon^2 \vg^2 T_y/2,
}\\ \displaystyle{(T_t,\genHx)=\genHx+\varepsilon\genGx-\varepsilon^2 T_x/2,
}\\ \displaystyle{(T_t,\genHy)=\genHy+\varepsilon\genGy-\varepsilon^2 T_y/2,
}\\ \displaystyle{(\genex,T_z)=T_z+\varepsilon T_x+\varepsilon^2 E_{0}^{-}/2,
}\\ \displaystyle{(\geney,T_z)=T_z+\varepsilon T_y+\varepsilon^2 E_{0}^{-}/2,
}\\ \displaystyle{(\genHx,T_z)=T_z+\varepsilon\vg^2\genex+\varepsilon^2\vg^2 E_{2}^{-}/4,
}\\ \displaystyle{(\genHy,T_z)=T_z+\varepsilon\vg^2\geney+\varepsilon^2\vg^2 E_{2}^{-}/4,
}\\ \displaystyle{(\genGx,T_t)=T_t+\varepsilon T_x+\varepsilon^2\vg E_{0}^{+}/2,
}\\ \displaystyle{(\genGy,T_t)=T_t+\varepsilon T_y+\varepsilon^2\vg E_{0}^{+}/2,
}\\ \displaystyle{(\genHx,T_t)=T_t-\varepsilon\genGx+\varepsilon^2\vg^2 E_{2}^{+}/4,
}\\ \displaystyle{(\genHy,T_t)=T_t-\varepsilon\genGy-\varepsilon^2\vg^2 E_{2}^{+}/4.
}
\end{array}
\label{TableAdjointOverflow444}
\end{eqnarray}

\end{landscape}

\tabcolsep=0.5truemm

\vspace{20pt}
\begin{table}[t]
\caption{344. Adjoint table for the Lie algebra (\ref{eq344Lie_algebra}).
The notation and convention are the same as in Table \ref{TableAdjoint466}.
$F_{0}^\pm= i k_0 \left[\gBone\pm\gBtwo\right]$.
\label{TableAdjoint344}
}
\begin{center}
\footnotesize
\begin{tabular}{|l|llllllllll|}
\hline &$T_x$ &$T_y$ &$T_z$ &$\Theta$ &$\gBone$ &$\gBtwo$ &$\genex$ &$\geney$ &$\gVthr$ &$\gWfou$
\\[5pt]\hline
$T_x$    &$.$ &$.$ &$.$ &$\Theta-\varepsilon T_y$ &$.$  &$.$ &$\genex-\varepsilon F_0^{-}$ &$.$
         &$\gVthr-\varepsilon T_x$ &$\gWfou-\varepsilon \genex+\varepsilon^2 F_0^{-}/2$
\\[5pt]
$T_y$    &$.$ &$.$ &$.$ &$\Theta+\varepsilon T_x$ &$.$  &$.$ &$.$ &$\geney-\varepsilon F_0^{-}$ 
         &$\gVthr-\varepsilon T_y$ &$\gWfou-\varepsilon \geney+\varepsilon^2 F_0^{-}/2$ 
\\[5pt]
$T_z$    &$.$ &$.$ &$.$ &$.$ &$.$ &$.$ 
         &$\genex-\varepsilon T_x$ &$\geney-\varepsilon T_y$  &$.$ &$.$ 
\\[5pt]
%
$\Theta$ &$T_x c+T_y s$ &$T_y c-T_x s$ &$.$ &$.$ &$.$ &$.$ 
      &$\genex c+\geney s$ &$\geney c-\genex s$ &$.$ &$.$ 
\\[5pt]
$\gBone$ &$.$ &$.$ &$.$ &$.$ &$.$ &$.$ &$.$ &$.$  &$.$ &$.$ 
\\[5pt]
$\gBtwo$ &$.$ &$.$ &$.$ &$.$ &$.$ &$.$ &$.$ &$.$  &$.$ &$.$ 
\\[5pt]
$\genex$ &$T_x+\varepsilon F_{0}^{-}$ &$.$ &$T_z+\varepsilon T_x+\varepsilon^2 F_{0}^{-}/2$ &$\Theta-\varepsilon\geney$ &$.$ 
      &$.$ &$.$ &$.$  &$.$ &$.$ 
\\[5pt]
$\geney$ &$T_x+\varepsilon F_{0}^{-}$ &$.$ &$T_z+\varepsilon T_y+\varepsilon^2 F_{0}^{-}/2$ &$\Theta-\varepsilon\geney$ &$.$ 
      &$.$ &$.$ &$.$  &$.$ &$.$ 
\\[5pt]
$\gVthr$&$T_x e^{\varepsilon}$ &$T_y e^{\varepsilon}$ &$T_z e^{2\varepsilon}$ &$.$ &$.$ &$.$ 
         &$\genex e^{-\varepsilon}$ &$\geney e^{-\varepsilon}$ &$.$ &$\gWfou e^{-2\varepsilon}$ 
\\[5pt]
$\gWfou$ &$T_x+\varepsilon\genex$ &$T_y+\varepsilon\geney$ &$T_z+\varepsilon\gVthr+\varepsilon^2\gWfou$ &$.$ 
         &$.$ &$.$ &$.$ &$.$ &$\gVthr+2\varepsilon\gWfou$ &$.$ 
\\[5pt]\hline
\end{tabular}
\end{center}
\end{table}

\vfill \eject

\section{Conclusion}

We have unveiled both the singularity structure and the underlying symmetries
of a nonlinear optics system which has great potential applications.
However, this analytic structure is too intricate 
to allow us to derive close form solutions.
Since the original nonlinear system results from a reductive perturbation method,
maybe another physical assumption during its derivation could
make it tractable by the analytic techniques investigated here.

\section*{Acknowledgments}

Both authors are happy to acknowledge the generous support of the 
Centro internacional de Ciencias in Cuernavaca.
RC is grateful to Universidad de C\'adiz for his visit,
and thanks the organizers of the Workshop on laser-matter interaction (WLMI, Porquerolles, 2012)
for their invitation.  


\vfill \eject 
\end{document}